\documentclass[12pt]{article}
\usepackage{graphicx}

\usepackage{amsmath}
\usepackage{amssymb}
\usepackage{axodraw}

\newcommand\pubnumber{DO--TH 15/13  \\  QFET--2015--26}
\newcommand\pubdate{\today}

\def\dortmund{Fakult\"at f\"ur Physik,\\
TU Dortmund, Otto-Hahn-Str.4, D-44221 Dortmund, Germany}

\def\Title#1{\begin{center} {\Large #1 } \end{center}}
\def\Author#1{\begin{center}{ \sc #1} \end{center}}
\def\Address#1{\begin{center}{ \it #1} \end{center}}

\newcommand\pubblock{\rightline{\begin{tabular}{l} \pubnumber\\
         \pubdate  \end{tabular}}}
\newenvironment{Abstract}{\begin{quotation}  }{\end{quotation}}
\newenvironment{Presented}{\begin{quotation} \begin{center} 
             PRESENTED AT\end{center}\bigskip 
      \begin{center}\begin{large}}{\end{large}\end{center} \end{quotation}}
\def\Acknowledgements{\bigskip  \bigskip \begin{center} \begin{large}
             \bf ACKNOWLEDGEMENTS \end{large}\end{center}}

\def\beq{\begin{equation}}
\def\eeq#1{\label{#1}\end{equation}}
\def\eeqn{\end{equation}}

\def\beqa{\begin{eqnarray}}
\def\eeqa#1{\label{#1}\end{eqnarray}}
\def\eeqan{\end{eqnarray}}

\let\bar=\overbar

\def\Dslash{\not{\hbox{\kern-4pt $D$}}}
\def\dslash{\not{\hbox{\kern-2pt $\del$}}}

\def\msb{{\bar{\ssstyle M \kern -1pt S}}}

\begin{document}
\begin{titlepage}
\pubblock

\vfill
\Title{Rare Semileptonic Charm Decays}
\vfill
\Author{Stefan de Boer}
\Address{\dortmund}
\vfill
\begin{Abstract}
An analysis of charm mesons decaying semileptonically via Flavor Changing Neutral Currents is presented.
We calculate the Wilson coefficients within the Standard Model.
A window in the decay distribution, where physics beyond the Standard Model could be measured is identified.
Exemplary, we study effects of leptoquark models.
\end{Abstract}
\vfill
\begin{Presented}
The 7th International Workshop on Charm Physics (CHARM 2015)\\
Detroit, MI, 18-22 May, 2015
\end{Presented}
\vfill
\end{titlepage}
\def\thefootnote{\fnsymbol{footnote}}
\setcounter{footnote}{0}
%

\section{Introduction}

\begin{figure}[htb]
\centering
 \begin{picture}(70,70)(0,0)
  \ArrowLine(10,60)(70,60)
  \ArrowLine(10,40)(30,30)
  \DashLine(30,30)(50,30){3}
  \ArrowLine(50,30)(70,40)
  \CArc(40,30)(10,180,360)
  \Photon(40,20)(50,10){1}{4}
  \ArrowLine(70,20)(50,10)
  \ArrowLine(50,10)(70,0)
  \Oval(10,50)(10,5)(0)
  \Text(10,50)[]{$D$}
  \Oval(70,50)(10,5)(0)
  \Text(70,50)[]{$P$}
  \Text(40,50)[b]{\small $\bar q$}
  \Text(20,40)[]{\small $c$}
  \Text(60,40)[]{\small $u$}
  \Text(40,10)[]{\small $\gamma$}
  \Text(70,25)[]{\small $\ell^+$}
  \Text(70,5)[b]{\small $\ell^-$}
  \Text(40,35)[]{\small $W$}
  \Text(25,20)[]{\small $q_d$}
 \end{picture}
 \caption{The $c\to u\ell\ell$ transition within the SM. The label $q_d$ denotes down-type quarks.}
 \label{fig:cull_sm}
\end{figure}
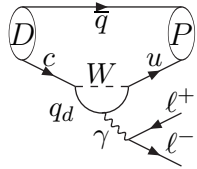

Rare semileptonic charm decays, e.g. the decay $D\to P\ell\ell$, where $D=c\bar q$, $q\in\{u,d\}$, $P$ is a pseudoscalar meson and $\ell$ is a muon or an electron are induced by $c\to u\ell\ell$ transitions.
Within the Standard Model (SM), the semileptonic decay of charm mesons via Flavor Changing Neutral Currents (FCNCs) is shown in Figure \ref{fig:cull_sm}.

In the SM, FCNCs are loop and Glashow-–Iliopoulos-–Maiani (GIM) suppressed.
The GIM suppression in $c\to u\ell\ell$ transitions is in particular effective due to the unitarity of the Cabibbo-–Kobayashi-–Maskawa (CKM) matrix $V$ and the small product $|V_{cb}^*V_{ub}|\approx1.5\cdot10^{-4}$.
Thus, $c\to u\ell\ell$ induced decays are rare in the SM and sensitive to Beyond Standard Model (BSM) physics as BSM dynamics could induce larger effective couplings and additional structures compared to the SM.
Rare semileptonic charm decays open windows to look for physics beyond the SM complementary to decays of $b$-quarks and $s$-quarks.
Additionally, $c\to u\ell\ell$ transitions probe perturbative QCD as the mass of the charm quark is close to the scale $\Lambda_\text{QCD}$.

Experiments on rare charm decays were performed and are designed by several collaborations, e.g. LHCb, CMS, BaBar, Belle (II), CLEO-c and BESIII.
The most stringent experimental limit to date is set on the mode $D^+\to\pi^+\mu^+\mu^-$ by the LHCb collaboration in 2013 reporting an upper limit on the fully integrated non-resonant branching fraction of \cite{Aaij:2013sua}
\begin{align}
 \mathcal B^{\text{exp}}(D^+\to\pi^+\mu^+\mu^-)<7.3\times10^{-8}\quad(90\%\,\text{CL})\,.
\end{align}
Additionally, upper limits in the low $q^2$ bin ($0.250\,\text{GeV}\le\sqrt{q^2}\le0.525\,\text{GeV}$) and in the high $q^2$ bin ($1.250\,\text{GeV}\le\sqrt{q^2}$), where $q^2$ is the dilepton mass squared read \cite{Aaij:2013sua}
\begin{align}
 &\mathcal B^{\text{exp,low}}(D^+\to\pi^+\mu^+\mu^-)<2.0\times10^{-8}\quad(90\%\,\text{CL})\,,\\
 &\mathcal B^{\text{exp,high}}(D^+\to\pi^+\mu^+\mu^-)<2.6\times10^{-8}\quad(90\%\,\text{CL})\,.
\end{align}

Calculations of the $c\to u\ell\ell$ branching fractions were performed by several groups.
Predictions for the fully integrated non-resonant SM branching fraction of $D^+\to\pi^+\mu^+\mu^-$ decays are given as
\begin{align}
 &\mathcal B_{D^+\to\pi^+\mu^+\mu^-}^{\text{nr,SM}}=6\cdot10^{-12}\quad\cite{Fajfer:2005ke}\,,\label{eq:Dtopill_Fajfer}\\
 &\mathcal B_{D^+\to\pi^+\mu^+\mu^-}^{\text{nr,SM}}=[4.59,8.04]\cdot10^{-10}\quad\cite{Wang:2014uiz}\,.\label{eq:Dtopill_Wang}
\end{align}
Equations (\ref{eq:Dtopill_Fajfer}) and (\ref{eq:Dtopill_Wang}) differ by two orders of magnitude.
This discrepancy persists in the predictions for the inclusive branching fractions
\begin{align}
 &\mathcal B_{D^+\to X_u^+e^+e^-}^{\text{nr,SM}}=2\cdot10^{-8}\quad\cite{Burdman:2001tf}\,,\label{eq:ctoull_Burdman}\\
 &\mathcal B_{D^+\to X_u^+e^+e^-}^{\text{nr,SM}}=6.0\cdot10^{-10}\quad\cite{Fajfer:2005ke}\label{eq:ctoull_Fajfer}\,,
\end{align}
whereas \cite{Paul:2011ar} gives a branching fraction consistent with equation (\ref{eq:ctoull_Burdman}).

Thus, we will do the following:
We calculate the SM branching fractions within the framework of an effective theory and identify the differences in the calculations of \cite{Wang:2014uiz}, \cite{Burdman:2001tf}, \cite{Paul:2011ar} and \cite{Fajfer:2005ke}.
Additionally, we look for measurable BSM effects in the decay distribution, e.g. due to leptoquark models.

\section{SM Branching Fractions}

In this section, we sketch the calculation of the SM branching fractions of the inclusive $c\to u\ell\ell$ decay and of the modes $D^+\to\pi^+\ell^+\ell^-$, where $\ell\in\{e,\mu\}$ \cite{deBoerSeidel}, \cite{deBoer:2015boa}.
By means of an Operator Product Expansion (OPE) we write the leading order effective weak Lagrangian at the weak scale as \cite{Greub:1996wn}
\begin{align}\label{eq:L_eff_weak}
 \mathcal L_\text{eff}^\text{weak}|_{\mu\sim\mu_W}=&{\frac{4G_F}{\sqrt 2}\sum_{q\in\{d,s,b\}}V_{cq}^*V_{uq}}\left(C_1(\mu)Q_1^{(q)}(\mu)+C_2(\mu)Q_2^{(q)}(\mu)\right)\,,
\end{align}
where $G_F$ denotes the Fermi constant.
The operators $Q_i^{(q)}$ read \cite{Chetyrkin:1996vx}
\begin{align}
 &Q_1^{(q)}=(\overline u_L\gamma_{\mu_1}T^aq_L)(\overline q_L\gamma^{\mu_1}T^ac_L)\,,\\
 &Q_2^{(q)}=(\overline u_L\gamma_{\mu_1}q_L)(\overline q_L\gamma^{\mu_1}c_L)\,,
\end{align}
where $T^a$ denotes the $SU(3)$ generators.
At the scale $\mu$, short-distance Wilson coefficients $C_i$ and the long-distance operators generated by light fields are factorized.

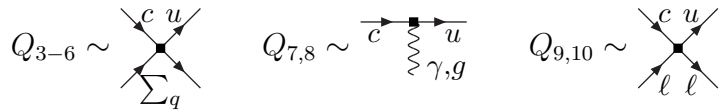
\begin{figure}[htb]
 \centering
 $Q_{3-6}\sim$
 \begin{picture}(30,27.5)(0,12.5)
  \ArrowLine(0,30)(15,15)
  \ArrowLine(15,15)(30,30)
  \ArrowLine(0,0)(15,15)
  \ArrowLine(15,15)(30,0)
  \CBoxc(15,15)(3,3){}{}
  \Text(10,25)[b]{\small $c$}
  \Text(20,25)[b]{\small $u$}
  \Text(15,5)[t]{\small $\sum_q$}
 \end{picture}\qquad
 $Q_{7,8}\sim$
 \begin{picture}(40,22.5)(0,7.5)
  \ArrowLine(0,20)(20,20)
  \ArrowLine(20,20)(40,20)
  \Photon(20,20)(20,0){2}{4}
  \CBoxc(20,20)(3,3){}{}
  \Text(5,15)[]{\small $c$}
  \Text(35,15)[]{\small $u$}
  \Text(25,5)[lt]{\small $\gamma$,$g$}
 \end{picture}\qquad
 $Q_{9,10}\sim$
 \begin{picture}(30,27.5)(0,12.5)
  \ArrowLine(0,30)(15,15)
  \ArrowLine(15,15)(30,30)
  \ArrowLine(0,0)(15,15)
  \ArrowLine(15,15)(30,0)
  \CBoxc(15,15)(3,3){}{}
  \Text(10,25)[b]{\small $c$}
  \Text(20,25)[b]{\small $u$}
  \Text(10,5)[t]{\small $\ell$}
  \Text(20,5)[t]{\small $\ell$}
 \end{picture}\\[1em]
 \caption{Pictorial representation of the additional operators due to the integration of the bottom quark.
 The label $q$ in the sum of $Q_{3-6}$ denotes light quark fields.}
 \label{fig:operators_at_mub}
\end{figure}

The Wilson coefficients are found by matching the SM Lagrangian onto the effective Lagrangian (\ref{eq:L_eff_weak}) at the weak scale.
Thus, heavy fields are integrated out, e.g. in QCD at Next-to-Next-to-Leading Order \cite{Bobeth:1999mk}.
The Wilson coefficients at lower scales are found via the solution of the Renormalization Group (RG) equation \cite{Gorbahn:2004my}.
At the threshold of the mass of the bottom quark it is integrated out generating the additional operators $Q_{3-10}$ shown in Figure \ref{fig:operators_at_mub}.
Thus, the Wilson coefficients $C_{1-10}(\mu_c)$ at the charm scale are found by resumming logarithms to all orders in perturbation theory by means of the RG.

We calculate the matrix elements of the operators $\langle Q_{1-6,8}\rangle$ in terms of effective Wilson coefficients.
The factorized hadronic matrix elements $\langle Q_{7,9,10}\rangle$ are parametrized via three form factors.
The form factors are related within the Heavy Quark Effective Theory yielding one independent form factor $f_+$.
We parametrize the form factor $f_+$ by means of the z-expansion, where the parameters are fitted to the experimentally measured $D\to\pi\ell\nu_\ell$ decay \cite{Amhis:2014hma}.

The calculation of the non-resonant SM branching fractions as sketched above yields \cite{deBoer:2015boa}
\begin{align}
 &\mathcal B_{D^+\to X_u^+e^+e^-}^\text{nr,SM}\approx1\cdot10^{-9}\,,\label{eq:BDXee}\\
 &\mathcal B_{D^+\to X_u^+\mu^+\mu^-}^\text{nr,SM}\approx2\cdot10^{-10}\,,\\
 &\mathcal B_{D^+\to\pi^+\ell^+\ell^-}^\text{nr,SM}\approx5\cdot10^{-12}\quad(\ell\in\{e,\mu\})\,.\label{eq:BDpill}
\end{align}
Our branching fractions (\ref{eq:BDXee})-(\ref{eq:BDpill}) are consistent with equations (\ref{eq:Dtopill_Fajfer}) and (\ref{eq:ctoull_Fajfer}) as calculated in \cite{Fajfer:2005ke}.
Compared to the calculations in \cite{Burdman:2001tf}, \cite{Paul:2011ar} and \cite{Wang:2014uiz} the primarily difference is due to the matching coefficients at the weak scale, e.g. \cite{Inami:1980fz}
\begin{align}
 C_9(\mu_W)=\sum_{q\in\{d,s,b\}}V_{cq}^*V_{uq}C_{9,\text{IL}}^{(q)}\approx V_{cs}^*V_{us}\frac{-2}9\ln\frac{m_s^2}{m_d^2}\approx-0.29\label{eq:C9_muW_IL}
\end{align}
which is vanishing in our matching.
By means of equation (\ref{eq:C9_muW_IL}) light quark fields are integrated out at the weak scale.
This, is not consistent within the OPE factorization of scales.
In particular, the logarithms as in $C_{9,\text{IL}}^{(q)}$ are resumed by means of the RG.

\section{BSM Physics and Leptoquarks}

\begin{figure}[htb]
 \centering
 \includegraphics[height=2.5in]{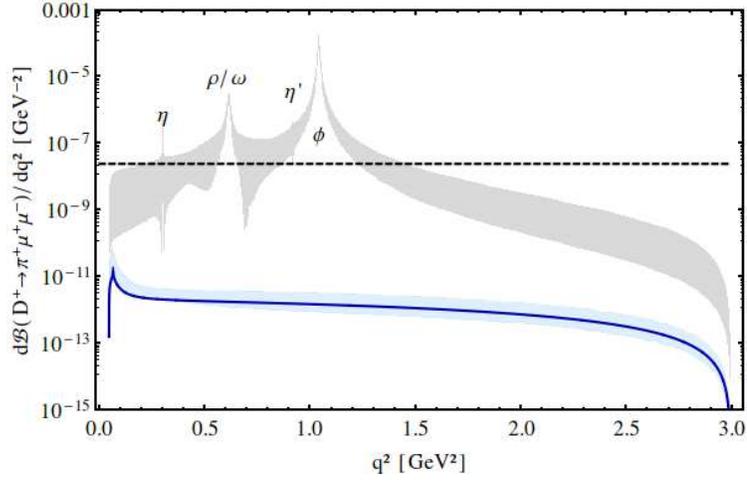}
 \caption{The $D^+\to\pi^+\mu^+\mu^-$ dilepton mass squared decay distribution.
 The solid blue curve is the non-resonant prediction at $\mu_c=\bar m_c$ and the light blue band its $\mu_c$-uncertainty.
 We find the gray resonant band by fitting the experimental branching fractions via a constant width Breit-Wigner shape and varying the relative strong resonant phases.
 The dashed black line is the binned non-resonant 90\% CL experimental upper limit taken from \cite{Aaij:2013sua}.
 Figure taken from \cite{deBoer:2015boa}.}
 \label{fig:dBQDppiR2mu}
\end{figure}

We add the resonant modes in the decay distribution by parameterizing them in terms of a Breit-Wigner shape, where the $\omega$ and $\rho$ resonances are related as deduced in \cite{Fajfer:2005ke} and we take the experimental upper limit for the $\eta'$ resonance.
The $D^+\to\pi^+\mu^+\mu^-$ decay distribution is shown in Figure \ref{fig:dBQDppiR2mu}.

In the SM, the branching fraction of the non-resonant mode is orders of magnitude smaller than the branching fraction of the resonant modes and likewise the current experimental upper limit.
Yet, at high $q^2$ the difference between the SM prediction and the experimental limit opens a window to look for BSM physics.
We predict the non-resonant SM branching fraction at high $q^2$ to be \cite{deBoer:2015boa}
\begin{align}
 \mathcal B_{D^+\to\pi^+\mu^+\mu^-}^\text{nr,SM}\big|_{q^2\geq1.525\,\text{GeV}^2}\approx7.4\cdot10^{-13}\;_{-14\%}^{+15\%}(\bar m_s)\,_{-45\%}^{+136\%}(\mu_c)\,_{-20\%}^{+27\%}(f_+)\,,
\end{align}
where uncertainties ($\bar m_c/\sqrt2\le\mu_c\le\sqrt2\bar m_c$) larger than ten percent are given, but we neglect power corrections.
The scale uncertainty is large due to a variation of $\mu_c$ close to $\Lambda_\text{QCD}$ and could be reduced via a calculation of the two-loop effective Wilson coefficient of $Q_9$ due to $\langle Q_{1,2}^{(d,s)}(\mu_c)\rangle$.

\begin{figure}[htb]
\centering
\includegraphics[height=2.5in]{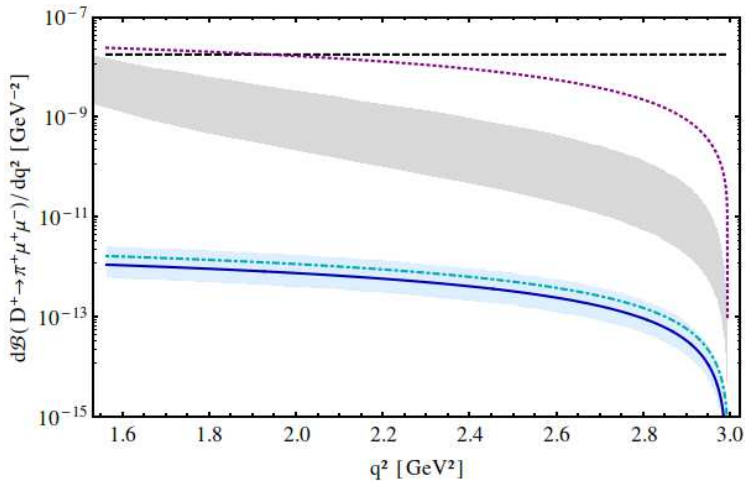}
\caption{The $D^+\to\pi^+\mu^+\mu^-$ dilepton mass squared decay distribution at high $q^2$.
 The solid blue curve is the non-resonant SM prediction at $\mu_c=\bar m_c$ and the light blue band its $\mu_c$-uncertainty, the gray band shows the resonant modes and the dashed black line denotes the 90\% CL experimental upper limit \cite{Aaij:2013sua}.
 Additionally, the dotted purple curve represents the vector leptoquark model and the dot-dashed cyan curve is the scalar leptoquark model.
 Figure adopted from \cite{deBoer:2015boa}.}
 \label{fig:dBQDppiR2muLQ}
\end{figure}

Clearly, any experimental signal in the branching fraction at high $q^2$ would be due to physics beyond the SM, e.g. a leptoquark (LQ) model.
Exemplary, we study effects of the low-energy scalar (3,3,-1/3) and vector (3,1,-5/3) LQ representations \cite{Davidson:1993qk}
\begin{align}
 &\lambda_{S_3}(\mathbf Q_L^Ti\tau_2\vec\tau\mathbf L_L)\cdot\vec{S_3}^\dagger\subset\mathcal L_{LQ}\,,\\
 &\lambda_{\tilde V_1}\bar q_R\gamma_\mu\ell_R(\tilde V_1^\mu)^\dagger\subset\mathcal L_{LQ}\,,
\end{align}
where $\mathbf Q$ are SM quark doublets, $\mathbf L$ are SM lepton doublets and $\tau_i$ denote the Pauli matrices.
Collider experiments constrain the LQ mass to be greater or similar than one TeV.
The products of couplings $\lambda$ inducing $c\to u\mu\mu$ transitions are constrained by the experimental limits on the branching fractions of $D^0\to\mu^+\mu^-$ and $D^+\to\pi^+\mu^+\mu^-$ decays.
As the scalar LQ model couples to quark doublets, its induced Wilson coefficients are additionally constrained by kaon decays.
For that purpose, we take the hierarchical flavor pattern of \cite{Varzielas:2015iva} to link the recent data on lepton flavor non-universality in rare semileptonic bottom decays.
The LQ induced decay distribution at high $q^2$ is shown in Figure \ref{fig:dBQDppiR2muLQ}.

At high $q^2$, the vector LQ model could induce branching fractions close to the experimental upper limit and the scalar LQ model induced branching fractions would degenerate into the SM prediction.

\section{Summary}
We have presented a calculation of the $c\to u\ell\ell$ Wilson coefficients within the SM.
The purpose of this calculation was to resolve discrepancies in the literature on the predictions for the non-resonant SM branching fractions.
Our fully integrated prediction $\mathcal B_{D^+\to\pi^+\mu^+\mu^-}^\text{nr,SM}\sim10^{-12}$ is orders of magnitude below the branching fraction of the resonant modes and likewise the current experimental upper limit $\mathcal B^\text{nr,exp}(D^+\to\pi^+\mu^+\mu^-)\lesssim10^{-8}$.
Yet, at high $q^2$ a window is identified, where BSM physics could be measurable.
Exemplary, we have studied effects of leptoquark models and found that a $SU(2)$-singlet vector leptoquark could induce branching fractions close to the current experimental limit.

\Acknowledgements
I would like to thank the organizers for the wonderful conference.
I am grateful to Gudrun Hiller, Bastian M\"uller and Dirk Seidel for a fruitful collaboration and Diganta Das for reading the manuscript.
This project is supported by the DFG Research Unit FOR 1873 ``Quark Flavour Physics and Effective Field Theories''.



\begin{thebibliography}{99}


\bibitem{Aaij:2013sua}
  R.~Aaij {\it et al.} [LHCb Collaboration],
  Phys.\ Lett.\ B {\bf 724} (2013) 203
  [arXiv:1304.6365 [hep-ex]].

\bibitem{Fajfer:2005ke}
  S.~Fajfer and S.~Prelovsek,
  Phys.\ Rev.\ D {\bf 73} (2006) 054026
  [hep-ph/0511048].

\bibitem{Wang:2014uiz}
  R.~M.~Wang, J.~H.~Sheng, J.~Zhu, Y.~Y.~Fan and Y.~G.~Xu,
  Int.\ J.\ Mod.\ Phys.\ A {\bf 30} (2015) 12,  1550063
  [arXiv:1409.0181 [hep-ph]].

\bibitem{Burdman:2001tf}
  G.~Burdman, E.~Golowich, J.~L.~Hewett and S.~Pakvasa,
  Phys.\ Rev.\ D {\bf 66} (2002) 014009
  [hep-ph/0112235].

\bibitem{Paul:2011ar}
  A.~Paul, I.~I.~Bigi and S.~Recksiegel,
  Phys.\ Rev.\ D {\bf 83} (2011) 114006
  [arXiv:1101.6053 [hep-ph]].

\bibitem{deBoerSeidel}
S.~de Boer, B.~M\"uller and D.~Seidel, to appear, DO-TH 15/11, QFET-2015-27

\bibitem{deBoer:2015boa} 
  S.~de Boer and G.~Hiller,
  arXiv:1510.00311 [hep-ph].

\bibitem{Greub:1996wn}
  C.~Greub, T.~Hurth, M.~Misiak and D.~Wyler,
  Phys.\ Lett.\ B {\bf 382} (1996) 415
  [hep-ph/9603417].
  
\bibitem{Chetyrkin:1996vx}
  K.~G.~Chetyrkin, M.~Misiak and M.~Munz,
  Phys.\ Lett.\ B {\bf 400} (1997) 206
   [Phys.\ Lett.\ B {\bf 425} (1998) 414]
  [hep-ph/9612313].

\bibitem{Bobeth:1999mk}
  C.~Bobeth, M.~Misiak and J.~Urban,
  Nucl.\ Phys.\ B {\bf 574} (2000) 291
  [hep-ph/9910220].

\bibitem{Gorbahn:2004my}
  M.~Gorbahn and U.~Haisch,
  Nucl.\ Phys.\ B {\bf 713} (2005) 291
  [hep-ph/0411071].

\bibitem{Amhis:2014hma}
  Y.~Amhis {\it et al.} [Heavy Flavor Averaging Group (HFAG) Collaboration],
  arXiv:1412.7515 [hep-ex].

\bibitem{Inami:1980fz}
  T.~Inami and C.~S.~Lim,
  Prog.\ Theor.\ Phys.\  {\bf 65} (1981) 297
   [Prog.\ Theor.\ Phys.\  {\bf 65} (1981) 1772].

\bibitem{Davidson:1993qk}
  S.~Davidson, D.~C.~Bailey and B.~A.~Campbell,
  Z.\ Phys.\ C {\bf 61} (1994) 613
  [hep-ph/9309310].

\bibitem{Varzielas:2015iva}
  I.~de Medeiros Varzielas and G.~Hiller,
  JHEP {\bf 1506} (2015) 072
  [arXiv:1503.01084 [hep-ph]].

\end{thebibliography}
\end{document}